\documentclass[12pt]{article}

\begin{document}

\title{An analytic model for the transition from  decelerated to accelerated cosmic expansion} 
\author{J. Ponce de Leon\thanks{E-mail: jponce@upracd.upr.clu.edu}\\ Laboratory of Theoretical Physics, Department of Physics\\ 
University of Puerto Rico, P.O. Box 23343, San Juan, \\ PR 00931, USA} 
\date{March  2006}

\maketitle

\begin{abstract}
We consider the scenario where our observable universe is devised as a dynamical four-dimensional  hypersurface embedded
 in a five-dimensional bulk spacetime, with a large extra dimension, which is the {\it generalization of the flat FRW cosmological metric to five dimensions}. This scenario generates a simple analytical model where different stages of the evolution of the universe are  approximated by distinct  parameterizations of the {\it same} spacetime. In this model 
the evolution from decelerated to accelerated expansion can be interpreted as  a   ``first-order" phase transition between two successive stages. The dominant energy condition allows different parts of the universe to evolve, from deceleration to acceleration, at different redshifts within a  narrow era. 
This picture corresponds to the creation of bubbles of  new phase, in the middle of the old one, typical of first-order phase transitions. Taking  $\Omega_{m} = 0.3$ today, we find that the cross-over from deceleration to acceleration occurs at $z \sim 1-1.5 $, regardless of the equation of state in the very early universe.
In the case of primordial radiation, the model predicts that  the deceleration parameter ``jumps" from $q \sim  + 1.5$ to $q \sim - 0.4$ at  $z \sim  1.17$. At the present time $q = - 0.55$ and the equation of state of the universe is $w  = p/\rho \sim - 0.7 $, in agreement with observations and some theoretical predictions.

\end{abstract}

\medskip



\newpage

\section{Introduction}
Observational data indicate  that, on large scales $( \gg 100)$Mps,  the universe is homogeneous and isotropic. Therefore,  in the standard model  
the 
geometry  of the universe  is described by the maximally-symmetric Friedmann-Robertson-Walker (FRW) line element. 
The simple power-law FRW models,  which correspond to  solutions of the Einstein field equations with the barotropic equation of state, have played an important role in cosmology explaining the overall features of the universe.   

However, these simple models are not adequate to describe the late evolution of the universe. Recent observations of the high red-shift of type $Ia$ supernovae \cite{Riess}-\cite{Tonry}, in concordance with data from measurements of the fluctuations in the power spectrum of the cosmic microwave background \cite{Lee}-\cite{Sievers},  indicate  that the expansion of the universe is speeding up rather than slowing down. Since radiation and ordinary matter are sources of attractive gravity, in general relativity an accelerating universe requires the existence of some repulsive energy, usually called ``dark" energy,  which dominates over matter and violates the strong energy condition $(\rho + 3p) > 0$.

On the other hand there is a substantial observational evidence that the accelerated expansion is a recent phenomenon. That is, dark energy was unimportant in the early evolution of the universe, but it is dominant during the late evolution \cite{Riess2},\cite{Turner3}. There are a number of models for dark energy that describe the transition from deceleration to acceleration, within the context of FRW cosmologies \cite{Peebles}-\cite{Krause}. 

Alternative explanations for the acceleration of the universe, beyond dark energy, include phantom energy \cite{Caldwell2}-\cite{Stefancic}, certain modifications of general relativity \cite{Turner1}-\cite{Easson}, the gravitational leakage into extra dimensions \cite{Lue}-\cite{Deffayet2}, Chaplygin gas \cite{Gorini}-\cite{Bento} as well as Cardassian models \cite{Freese}-\cite{Gondolo}. In braneworld theory, non-static extra dimensions can generally lead to a variable vacuum energy, which in turn may explain the present accelerated cosmic expansion \cite{JPdeL0305041}-\cite{JPdeL0412005}.

In this paper we present a simple model where the recent accelerated cosmic expansion is a natural consequence of embedding our universe, as a dynamical four-dimensional  hypersurface, in a five-dimensional bulk spacetime, with a large extra dimension, which is the generalization of  the flat FRW cosmological metric in $4D$. 
 
Let us explain the details of the model. In four-dimensional general relativity, when the content of the  universe is taken to be a perfect fluid satisfying the barotropic equation of state, the geometry of the spatially flat FRW universe is described, in the usual coordinates,  by the metric  
\begin{equation}
\label{FRW spatially flat model}
ds^2 = dt^2 - A^2 t^{2/\alpha}[dr^2 + r^2(d\theta^2 + \sin^2\theta d\phi^2)],
\end{equation}
where $\alpha$ is a dimensionless free parameter, and $A$ is an arbitrary constant with dimension $L^{- 1/\alpha}$. The corresponding energy density $\rho$ and pressure $p$ of the cosmological fluid are given by  
\begin{equation}
\label{density on y = const}
8 \pi G \rho = \frac{3}{\alpha^2 t^2},\;\;\;\mbox{and}\;\;\;p = \gamma \rho,\;\;\;\gamma = \frac{(2\alpha - 3)}{3}.
\end{equation}

The generalization of (\ref{FRW spatially flat model}) to five dimensions is given by the line element 
\begin{equation}
\label{Ponce de Leon solution}
d{\cal S}^2 = y^2 dt^2 - A^2 t^{2/\alpha}y^{2/(1 - \alpha)}[dr^2 + r^2(d\theta^2 + \sin^2\theta d\phi^2)] - \alpha^2(1- \alpha)^{-2} t^2 dy^2,
\end{equation}
where the unitless ``fifth" coordinate $y$ represents the large spacelike ``extra" dimension. 
This is a solution of the five-dimensional Einstein equations in $5D$ vacuum \cite{JPdeL 1}. 
It has been applied to the discussion of a wide variety of cosmological problems  \cite{Overduin}-\cite{equ. STM-Brane}, and in the literature of space-time-matter theory (STM) is known as the {\it standard} cosmological model in $5D$ \cite{Wesson book}.

In four dimensions, the line element (\ref{Ponce de Leon solution})  reduces to the familiar FRW metric (\ref{FRW spatially flat model}) on any hypersurface $\Sigma_{0}: y = y_{0}$ orthogonal to the extra dimension, and the induced energy-momentum tensor gives back the expressions for density and pressure (\ref{density on y = const}).

However, this embedding is not unique; there is more than one way for embedding  a $4D$ spacetime in a given five-dimensional manifold.
Indeed, on any other four-dimensional hypersurface $\Sigma: y = f(t)$, which is non-orthogonal to $y$-lines, the induced metric is \cite{JPdeL/0511067}
\begin{equation}
ds^2_{\Sigma} = \left(f^2 - \frac{\alpha^2}{(1 - \alpha)^2}t^2 \dot{f}^2 \right)dt^2 - A^2 t^{2/\alpha}y^{2/(1 - \alpha)}[dr^2 + r^2(d\theta^2 + \sin^2\theta d\phi^2)],
\end{equation}
where $\dot{f} = df/dt$. On $\Sigma$ we wish to  recover the flat FRW cosmological metric
  \begin{equation}
\label{FRW from the brane}
ds^2_{\Sigma} = dt^2 - a^2(t)\left[ dr^2 + r^2 d\Omega^2\right].
\end{equation}
Consequently, continuity requirements across $\Sigma$ demand  
\begin{equation}
\label{defining f}
f^2 - \frac{\alpha^2}{(1 - \alpha)^2}t^2 \dot{f}^2 = 1,
\end{equation}
and 
\begin{equation}
\label{defining a}
a(t) = At^{1/\alpha}f^{1/(1 - \alpha)}.
\end{equation}
Equation (\ref{defining f}) has two solutions: $f = 1$ and 
\begin{equation}
\label{solution for f}
f(t) = \frac{1}{2}\left\{C t^{(1 - \alpha)/\alpha} + \frac{1}{C t^{(1 - \alpha)/\alpha}}\right\},
\end{equation}
where $C$ is a constant of integration with units of $L^{(\alpha - 1)/\alpha}$. The first solution reproduces the usual FRW scenario (\ref{FRW spatially flat model}) in $4D$, while  the second one is similar to the power-law FRW models only for ``small" and ``large" times. Consequently,  on the dynamical hypersurface $\Sigma: y = f(t)$ two distinct $4D$ scenarios emerge from  (\ref{defining a}) and (\ref{solution for f}), depending on whether $\alpha > 1$ or $\alpha < 1$. After some manipulations, we get   
\begin{equation}
\label{original a1}
ds_{1}^2 = dt^2 -  \frac{A^{2}_{1} t^{4/\alpha}}{\left[C_{1}^2 + t^{2(\alpha - 1)/\alpha}\right]^{2/(\alpha - 1)}}\left[dr^2 + r^2 (d\theta^2 + \sin^2\theta d\phi^2 )^2\right],\;\;\;\alpha > 1
\end{equation}
and\footnote{In what follows, in order to avoid any confusion we will use $\beta$ to denote $\alpha < 1$} 
\begin{equation}
\label{original a2}
ds_{2}^2 = dt^2 -  A^2_{2}\left[1 + C_{2}^2t^{2(1 - \beta)/\beta} \right]^{2/(1 - \beta)}\left[dr^2 + r^2 (d\theta^2 + \sin^2\theta d\phi^2 )^2\right],\;\;\;\beta < 1.
\end{equation}
where $A_{1}$, $A_{2}$, $C_{1}$ as well as $C_{2}$ are constants with the appropriate dimensions, while   $\alpha$ and $\beta$ are dimensionless parameters. These two metrics constitute the starting point for our $4D$ model. 
 
We show that the first metric
describes the early (post inflationary) evolution of the universe,
 while  the second one describes the late (post recombination)
evolution of the universe. We require the fulfillment of physical conditions, as well as the continuity of the first and second
fundamental forms, across the appropriate $3D$ surface, in order to show that the metrics (\ref{original a1}) and  (\ref{original a2}) can be considered as representing consecutive stages in the evolution of the universe. 

Thus,  on the dynamical spacetime section $\Sigma: y = f(t)$, the  {\it standard} cosmological model in $5D$ (\ref{Ponce de Leon solution}) yields  a simple model in $4D$  where different stages of the evolution of our flat FRW universe are  approximated by distinct  parameterizations of the {\it same} spacetime. Indeed, by changing $\alpha \longleftrightarrow \beta$ and redefining the constants we see that $(\ref{original a1}) \longleftrightarrow (\ref{original a2})$. Once again we see that apparent different scenarios in $4D$ may correspond to the same physics in  $5D$ \cite{JPdeLgr-qc/0512067}.

Throughout the paper we assume that all matter fields and fluids,  have positive energy density and satisfy the dominant energy condition $- p \leq \rho \leq p$.
Under these terms we obtain several interesting results. 
\begin{enumerate}
\item The deceleration parameter today is $q = - 1 + 3\Omega_{m}/2$, which for $\Omega _{m} = 0.3$ gives $q = - 0.55$, in agreement with observations, \cite{Moncy}-\cite{DD1}.
\item For the equation of state of the universe $w = p/\rho$,  we obtain $w = (p/\rho) = -1 + \Omega_{m}$. That is,  $w = - 0.7$ for $\Omega _{m} = 0.3$ today, in accordance with observations and other theoretical models \cite{Alam}-\cite{Steinhard}.
\item In our model the cross-over from decelerated to accelerated cosmic expansion is a   ``first-order" phase transition where the pressure is discontinuous and  involves the release 
of ``latent" energy accumulated during the early evolution of the universe. The transition corresponds to a discontinuity in a Higgs-type scalar field.
\item The transition  is ``almost" insensitive to the equation of state in the very early universe; it begins at approximately the same redshift era, viz., $z \sim 1-1.5$, for a wide variety of primordial matter. In the case of primordial radiation, and $\Omega_{m} = 0.3$ today,  the transition occurs at  $z = 1.16 \pm 0.12$.
\item The dominant energy condition allows different parts of the universe to evolve, from one phase to another, at different redshifts within a  narrow era. 
This picture corresponds to the creation of bubbles of  new phase, in the middle of the old one, typical of first-order phase transitions. Only for primordial stiff matter, the transition is an instantaneous phenomenon that occurs everywhere at the same time.
\item The duration of the transition   
depend on how ``stiff",  or ``soft"  is the primordial matter; the softer the equation of state,  the longer the transition.
\end{enumerate}
This paper is organized as follows. In section $2$ we show that   metric (\ref{original a1}) can indeed be used to describe an early evolution of the universe. 
In section $3$, we discuss the physical consequences of metric (\ref{original a2}) in terms of the  density and deceleration parameters.  We use them to constraint $\beta$ as well as to  obtain the age of the universe. 
In section $4$, we show that, without contradicting physical requirements, the cross-over from decelerated to accelerated expansion can be depicted as  a phase transition of first-order from the epoch described by (\ref{original a1}) to the one described by (\ref{original a2}). 
In section $5$ we present an estimate for $\beta$ and study the ``thin" era sandwiched between, the end of the early evolution described by  (\ref{original a1}), and the beginning of the late evolution of the universe, as described by (\ref{original a2}). 
 In section $6$ we discuss the results.

\section{Model for the early evolution of the universe} 
In this section we study the features of the line element given by (\ref{original a1}). We show that it can be used to describe an early evolution of the universe. Then we use the dominant energy condition to put a  ceiling on its applicability.

The cosmic scale factor is given by
\begin{equation}
\label{a1}
a_{1}(t) = \frac{A_{1} t^{2/\alpha}}{\left[C_{1}^2 + t^{2(\alpha - 1)/\alpha}\right]^{1/(\alpha - 1)}}, 
\end{equation}
with 
\begin{equation}
\alpha > 1.
\end{equation}
In what follows, in order to simplify the notation, we introduce the dimensionless coordinate $x$ as
\begin{equation}
\label{x}
x = C_{1}^{- 2}t^{2(\alpha - 1)/\alpha}. 
\end{equation}
With this notation, the cosmic factor (\ref{a1}) becomes
\begin{equation}
\label{a1 in terms of x}
a_{1}(t) = A_{1}\left[\frac{x}{1 + x}\right]^{1/(\alpha - 1)}.
\end{equation}
So, $A_{1}$ is a dimensionless constant, and $C_{1}$ has dimensions $L^{(\alpha - 1)/\alpha}$.

The kinematical quantities to be used in our discussion are, the Hubble parameter 
\begin{equation}
H_{1} = \frac{2}{\alpha t(1 + x)},
\end{equation}
and the deceleration parameter 
\begin{equation}
\label{deceleration parameter}
q_{1}  = - \frac{\ddot{a}a}{{\dot{a}}^2} = \frac{1}{2}\left[\alpha - 2 + (3\alpha - 2)x\right].
\end{equation}
The matter density and pressure are obtained from the Einstein field equations. We get
\begin{equation}
8 \pi G \rho_{1} = \frac{12}{\alpha^2 t^2(1 + x)^2},
\end{equation}
and 
\begin{equation}
8\pi G p_{1} = \frac{4(\alpha - 3) + 4(3\alpha - 2)x}{\alpha^2 t^2(1 + x)^2}. 
\end{equation} 
In the very early universe,  the model exhibits  good physical properties. Namely, for ``small" values of $t$, it  reduces to $a_{1} \approx t^{2/\alpha}$, and the corresponding primordial matter distribution is
\begin{equation}
\label{matter quantities for the very early universe}
8\pi G\rho = \frac{12}{\alpha^2 t^2}, \;\;\;8\pi Gp = \frac{4(\alpha - 3)}{\alpha^2t^2},\;\;\;\frac{p}{\rho} = \frac{\alpha - 3}{3}.
\end{equation}
The deceleration parameter (\ref{deceleration parameter}) reduces to 
\begin{equation}
\label{q asymptotically}
q =  \frac{1}{2}(\alpha - 2),
\end{equation}
and the gravitational density of the induced matter is 
\begin{equation}
\label{gravitationa density for primordial matter}
(\rho + 3p) = (\alpha - 2)\rho. 
\end{equation}
The above shows that (i) for $\alpha > 2$ the primordial matter behaves similar to ordinary gravitating matter (like dust for $\alpha = 3$, radiation  for $\alpha = 4$, or stiff matter for $\alpha = 6$) where  the expansion is slowing down; (ii) for $\alpha = 2$ the  primordial matter behaves like a network of cosmic strings $(p= - \rho/3)$ and (iii)   
for $1 < \alpha < 2$  the primordial cosmological ``fluid" has repulsive properties; it violates the strong energy condition $(\rho + 3p) > 0$ and the deceleration parameter is negative. 

At late times, the metric $a_{1}$ shows two ``exotic" features.
First, instead of expanding to infinity, $a_{1}$ moves toward a constant finite value, viz., $a_{1}(t) \rightarrow A_{1}$, for large  $t$. In other words,  $a_{1}$ is  a spatially flat  coasting cosmology, approaching Minkowski space, asymptotically. In terms of the redshift $z$;  the coordinate $t$ used in $a_{1}$ does not cover the whole range $z \;\;\epsilon \;\;(0, \infty)$; where $z = 0$ and $z = \infty$ represent the redshift today and at the big bang, respectively.

Second, the ratio 
\begin{equation}
\frac{p_{1}}{\rho_{1}} = \frac{1}{3}[(\alpha - 3) + (3\alpha - 2)x],
\end{equation}
increases to infinity with time. 

In order to get rid of these  exotic features, we impose some physical conditions. In this paper we assume that all matter fields and fluids,  have positive energy density and satisfy the dominant energy condition $- p \leq \rho \leq p$.

For the case under consideration, this condition yields an upper limit  for $x$. Namely,
\begin{equation}
\label{dominant energy condition}
x_{max} = \frac{6 - \alpha}{3\alpha - 2}.
\end{equation}
Thus, the solution above will be restricted to
\begin{equation}
\label{restriction on x}
 0 \leq x \leq x_{max}.
\end{equation}
The dominant energy condition applied to primordial matter confines the possible values of  $\alpha$ in $a_{1}(t)$. Namely,  
\begin{equation}
\label{range of alpha}
1 < \alpha \leq 6.
\end{equation}
We will return to this point, with explicit examples, in section $5$.

\section{Model for the late evolution of the universe}
We now proceed to show that the line element (\ref{original a2}) can be used to describe the late evolution of the universe. We use the dominant energy condition to put a lower limit on its applicability. Then, we discuss the interpretation of the matter quantities and find the appropriate expressions for the deceleration and density parameters.
 
 The cosmic scale factor is given by
\begin{equation}
\label{a2}
a_{2}(t) = A_{2}\left[1 + C_{2}^2t^{2(1 - \beta)/\beta} \right]^{1/(1 - \beta)}, 
\end{equation}
with 
\begin{equation}
 \beta < 1, \;\;\;\beta \neq 0.
\end{equation}
In this case the dimensionless coordinate, which now we denote by $y$, is
\begin{equation}
\label{y}
y = C_{2}^2 t^{2{(1 - \beta)/\beta}}.
\end{equation}
 In terms of this coordinate, the scale factor becomes
\begin{equation}
\label{ay}
a_{2}(t) = A_{2}\left[1 + y\right]^{1/(1 - \beta)}.
\end{equation}
The corresponding kinematical quantities are given by,
\begin{eqnarray}
\label{kinematical quantities for y-solution}
H_{2} &=& \frac{2y}{\beta t(1 + y)},\nonumber \\
q_{2} &=& \frac{1}{2}(\beta - 2) + \frac{(3\beta -2)}{2 y}.
\end{eqnarray}
The matter density and pressure are
\begin{eqnarray}
\label{effective quantities}
 8 \pi G \rho_{2} &=& \frac{12y^2}{\beta^2 t^2(1 + y)^2},\nonumber \\
8\pi G p_{2} &=& \frac{4(\beta - 3)y^2 + 4(3\beta - 2)y}{\beta^2 t^2(1 + y)^2}. 
\end{eqnarray}
This model exhibits the same exotic features as the previous one, but for $t \rightarrow 0$, i.e., for  $y \approx 0$. First, $a_{2} = A_{2}$ at $y = 0$, instead of the usual $a(0) = 0$ in FRW models. Second, $(p_{2}/\rho_{2}) \rightarrow \infty$ for $y \rightarrow 0$. 

Again, the dominant energy condition puts away these unwanted features. This time it yields  a lower limit for $y$. Namely, 
\begin{equation}
\label{ymin}
y_{min} = \frac{3\beta - 2}{6 - \beta}.
\end{equation}
Thus, for the model under consideration we require 
\begin{equation}
y > y_{min}.
\end{equation}
For large values of $t$ the model exhibits features that are consistent with new results in observational cosmology.  To begin with, we notice that the deceleration parameter $q_{2}$ becomes negative, meaning that the late expansion of the universe is speeding up. Evidence in favor of a recent accelerated expansion is provided by  observations of high-redshift supernovae Ia \cite{Riess}-\cite{Tonry}, as well as  other observations, including  the cosmic microwave background and galaxy power  spectra \cite{Lee}-\cite{Sievers}.

On the other hand,  for large values of $t$ (or $y$), the pressure is negative. In order to interpret this, the usual approach is to invoke the existence of some kind of matter, sometimes  called missing  or dark energy, which possesses a large negative pressure \cite{Pelmutter2}. The standard candidate for this  energy is the cosmological constant \cite{Riess}-\cite{Padmanabhan0}, which looks like an ideal fluid with negative pressure  $p_{\Lambda} = - \rho_{\Lambda}$.

Accordingly, we assume that the universe is filled with ordinary matter and a cosmological term, i.e., 
\begin{eqnarray}
\label{splitting the effective quantities}
8\pi G\rho &=& 8\pi G\rho_{m}  + \Lambda,\nonumber \\ 
8\pi Gp &=& 8\pi Gp_{m}  - \Lambda.
\end{eqnarray}
Since the overwhelming time of the evolution of the universe is spent in the matter-dominated domain we  set $p_{m} = 0$. Thus, we obtain 
\begin{eqnarray}
\label{density for the regular matter}
8 \pi G \rho_{m} = \frac{4\beta y^2 + 4(3\beta - 2)y}{\beta^2 t^2(1 + y)^2},
\end{eqnarray}
and
\begin{equation}
\label{cosmological term}
\Lambda = 4 \frac{(3 - \beta)y^2 -(3\beta - 2)y}{\beta^2 t^2(1 + y)^2}.
\end{equation}
We note that, although the energy-momentum tensor for vacuum looks like an ideal fluid with negative pressure, the dynamics of a varying vacuum {\it cannot} be that of a fluid because stability would require $dp/d\rho >0$. 
The simplest microphysical model for a variable $\Lambda$, as well as for quintessence,  is the 
dark energy of a single scalar field $\phi$ with Lagrangian density given by
\begin{equation}
{\cal{L}} = - \frac{1}{2}\partial_{\mu}\phi\partial^{\mu} \phi - V(\phi),
\end{equation} 
where $V(\phi)$ is a self-interaction potential. The stress-energy tensor for this field
\begin{equation}
T_{\mu\nu} = -\partial_{\mu} \phi\partial_{\nu} \phi - {\cal{L}}g_{\mu\nu},
\end{equation}
looks like a perfect fluid with energy density and pressure given by 
\begin{eqnarray}
\rho_{\phi} &=& \frac{1}{2}{\dot{\phi}}^2 + V(\phi),\nonumber \\
p_{\phi} &=& \frac{1}{2}{\dot{\phi}}^2 - V(\phi),
\end{eqnarray}
where we have neglected the spatial gradient of $\phi$.

In this scenario, when the scalar field $\phi$ evolves very slowly and ${\dot{\phi}}^2 \ll V$, the field energy approximates the effect of a cosmological term with
\begin{equation} 
\Lambda = \rho_{\phi} \approx - p_{\phi}, 
\end{equation}
and the universe is dominated by the vacuum energy, which is responsible for the observed present acceleration.  

\subsection{The density parameter}

An essential  quantity in cosmology is the so called  density parameter $\Omega_{m}$, which is  
\begin{equation}
\label{density parameter}
\Omega_{m} = \frac{8 \pi G \rho_{m}}{3H^3}= \frac{1}{3}\left[{\beta +(3\beta - 2)y^{- 1}}\right].
\end{equation}
This quantity is an observable and is related to other important observational parameters. 
\paragraph{Relation between $q$ and $\Omega_{m}$:} Substituting (\ref{density parameter}) into the deceleration parameter given by (\ref{kinematical quantities for y-solution}), we get
\begin{equation}
\label{general formula for the acceleration}
q_{2} = -1 + \frac{3}{2} \Omega_{m},
\end{equation}
regardless of $\beta$. Thus, the deceleration parameter can be determined by measurements of $\bar{\Omega}_{m}$, and vice-versa.  In particular, if we take ${\bar{\Omega}}_{m} = 0.3$ today, then we obtain
\begin{equation}
\bar{q}_{2} = - 0.55,
\end{equation}
for the present acceleration, which is within the region of the suspected values for the present acceleration of the universe \cite{Moncy}-\cite{DD1}. 

\paragraph{Equation of state of the universe:} For the total, or effective,  energy density and pressure the equation of state of the universe can be written as
\begin{equation}
\label{definition of w}
w = \frac{p}{\rho}.
\end{equation}
Then, from (\ref{effective quantities}) and (\ref{density parameter}) we find
\begin{equation}
\label{equation of state for the universe}
w = -1 + \Omega_{m},
\end{equation}
which for ${\bar{\Omega}}_{m} = 0.3$ gives $w = - 0.7$ in  
agreement with observations \cite{Alam}-\cite{Steinhard}. 

\paragraph{Restrictions on $\beta$:} The parameter $\beta$ has thus far only one condition, viz., $2/3 < \beta < 6$, which follows from (\ref{ymin}). A more rigorous constraint comes from the density parameter. Evaluating 
(\ref{density parameter}) today and solving for $\bar{y}$ we find 
\begin{equation}
\label{condition for beta}
\bar{y} = C_{2}^2 \bar{t}^{2{(1 - \beta)/\beta}} = \frac{(3\beta - 2)}{(3\bar{\Omega}_{m} - \beta)},
\end{equation} 
where the $\bar{t}$ and $\bar{\Omega}_{m}$ represent the present value of $t$ and $\Omega_{m}$, respectively. 
Then, from the positivity of $y$ we find
\begin{equation}
\label{range of beta}
\frac{2}{3} < \beta < 3 \bar{\Omega}_{m}.
\end{equation}
In particular,  for $\bar{\Omega}_{m} = 0.3$ today we have that $\beta$ can take any value between $2/3$ and $0.9$. We note that if $\beta = 2/3$, then from (\ref{density parameter}) it follows that $\Omega_{m} = \beta/3$, i.e, $\Omega_{m} = 2/9 \approx 0.22$, so that $\bar{y}$ remains finite. 

\paragraph{Age of the universe:} In order to obtain the age of the universe, $\bar{t}$,  we evaluate the  Hubble parameter from (\ref{kinematical quantities for y-solution})  and use the above expression for $\bar{y}$. We  obtain  
\begin{equation}
\label{age of the universe}
\bar{t} = {F(\bar{\Omega}_{m}, \beta )}{\bar{H}}^{- 1},
\end{equation}
where
\begin{equation}
\label{F}
F(\bar{\Omega}_{m}, \beta ) = \frac{2(3\beta - 2)}{\beta (3 {\bar{\Omega}}_{m} + 2\beta -2)}.
\end{equation}
Here, and it what follows, $\bar{H} = {\bar{H}}_{2}$ is the present value of the Hubble ``constant". 

We see that  the age of the universe depends on two parameters; viz., $\bar{\Omega}_{m}$ and $\beta$. This is different from the familiar FRW cosmologies where $\bar{t}$ is completely determined by $\Omega_{m}$.  Note that  $F$ is an increasing function of $\beta$. Therefore, for a fixed  $\bar{\Omega}_{m} = 0.3$, we find that the age of the universe is larger than the dust FRW model for any $\beta > 0.688$.

\section{The interface between early and late evolution}  

In this section we prove that the  metrics (\ref{original a1}) and (\ref{original a2}) can indeed be considered as representing two consecutive stages in the evolution of the universe. We  show that the junction conditions can be satisfied without contradicting   physical requirements.
We also present the formulae relating the dimensionless coordinates $x$ and $y$ to the redshift $z$, which is the directly observed quantity used in research of distant objects. 

\subsection{Matching conditions}

We now proceed to match the metrics (\ref{original a1}) and (\ref{original a2}) across a $3D$ surface $t = \stackrel{\ast}{t} = Const$. At this surface we require continuity of the cosmic factor and its first derivative, viz., $a_{1}(\stackrel{\ast}{x}) = a_{2}(\stackrel{\ast}{y})$ and ${\dot{a}}_{1}(\stackrel{\ast}{x}) = {\dot{a}}_{2}(\stackrel{\ast}{y})$. This assures the continuity of the mass function, and the Hubble term, for interior and exterior geometry. 

It should be mentioned that we {\it cannot} require continuity of the second derivative of the metric; this would lead to negative values for $\stackrel{\ast}{x}$. As a consequence, in or model, the cross-over from decelerated to accelerated cosmic expansion is a   first-order phase transition where the pressure is discontinuous\footnote{First-order phase transitions exhibit a discontinuity in the first derivative of the free energy $F = F(V, T)$.  
The pressure is $p = - (\partial F/\partial V)_{|T}$, where $V$ and $T$ are the volume and temperature, respectively.}.    

Thus, the equations to be solved are 
\begin{eqnarray}
\label{boundary conditions}
A_{1}\left[\frac{\stackrel{\ast}{x}}{1 + \stackrel{\ast}{x}}\right]^{1/(\alpha - 1)} &=&  A_{2}[{1 + \stackrel{\ast}{y}}]^{1/(1 - \beta)}, \nonumber \\
\frac{1}{\alpha(1 + \stackrel{\ast}{x})} &=& \frac{ \stackrel{\ast}{y}}{\beta (1 + \stackrel{\ast}{y})}.
\end{eqnarray}
From the second equation we find
\begin{equation}
\label{x from the boundary conditions}
\stackrel{\ast}{x} = \frac{[\beta - (\alpha - \beta)\stackrel{\ast}{y}]}{\alpha \stackrel{\ast}{y}}.
\end{equation}
Substituting this into the first one, we get
\begin{equation}
\label{equation for eta}
(1 + \stackrel{\ast}{y})^{[(\alpha - \beta)/(1 - \beta)]} + \frac{\eta(\alpha - \beta)}{\beta}(1 + \stackrel{\ast}{y}) - \frac{\eta\alpha}{\beta} = 0,
\end{equation} 
where $\eta$ denotes the dimensionless constant
\begin{equation}
\label{eta}
\eta = \left(\frac{A_{1}}{A_{2}}\right)^{(\alpha - 1)}. 
\end{equation}
Thus, for any given $\alpha$ and $\beta$ we have two equations for three unknown, viz., $\stackrel{\ast}{x}$, $\stackrel{\ast}{y}$ and $\eta$. Obviously, there are infinite solutions to these equations. Thus, in order to restrict them we should impose some physical conditions, in addition to the evident requirement $\stackrel{\ast}{x} > 0$, $\stackrel{\ast}{y} >0$ and $\eta > 0$. 

\subsection{The dominant energy condition}

Note that, for the late epoch described by $a_{2}$,  the dominant energy condition  is automatically satisfied  by virtue of (\ref{equation of state for the universe}),  so that it needs not to be considered as a constraint. Now, the positivity of $y$ requires (\ref{range of beta}), which in turn guarantees the positivity of the matter density (\ref{density for the regular matter}).

The dominant energy condition in the epoch described by $a_{1}$ sets a lower limit on $\stackrel{\ast}{y}$, while $\stackrel{\ast}{x} > 0$ gives an upper bound. Namely,  
from (\ref{dominant energy condition}) and (\ref {x from the boundary conditions}), we find 
\begin{equation}
\label{more stringent constraint on y}
\frac{\beta(3\alpha - 2)}{[2\alpha(2 + \alpha) - \beta(3\alpha - 2)]} < \; \stackrel{\ast}{y}\; < \frac{\beta}{\alpha - \beta}.
\end{equation}
It should be mentioned that the denominator here is positive in the whole range of values for $\alpha$ and  $\beta$.
We note that (\ref{more stringent constraint on y}) subsumes condition (\ref{ymin}), for all $\alpha$. It also subsumes condition $\stackrel{\ast}{y} > (3\beta - 2)/(3 - \beta)$ required for the  positivity of $\Lambda$ in (\ref{cosmological term}), except for models where  $\beta$ is close to its maximum value $\beta \approx 3\bar{\Omega}_{m}$. But, these models are excluded by recent cosmological data, as we will see bellow.

Next, from (\ref{equation for eta}) we obtain
\begin{equation}
\label{positivity of eta}
\eta = \frac{\beta(1 + \stackrel{\ast}{y})^{[(\alpha - \beta)/(1 - \beta)]}}{(\alpha - \beta)}
\left[\frac{\beta}{(\alpha - \beta)} - \stackrel{\ast}{y}\right]^{- 1}.
\end{equation}
Thus, by virtue of (\ref{more stringent constraint on y}), $\eta$ is always positive as required. 

In summary,   for every $\alpha$ and $\beta$, the dominant energy condition provides     the range of ${\stackrel{\ast}{x}}$ and ${\stackrel{\ast}{y}}$ for which the metrics (\ref{original a1}) and (\ref{original a2}) can be matched. Therefore, without contradicting physical requirements, the cross-over from decelerated to accelerated expansion can be depicted as  a phase transition of first-order. 

\subsection{The redshift $z$}

The description of the evolution of the universe in terms of $x$ and $y$ is impractical because they  are not observed quantities. Our purpose now is to express them  in terms of  $z$, the redshift of the light emitted at time $t$, which is the directly observed quantity used in research on distant objects. 

The connection between the scale factor and the redshift parameter is given by 
\begin{equation}
\label{definition of z}
a(t) = \frac{a(\bar{t})}{1 + z}.
\end{equation}
At the present time $z = 0$. Thus, using (\ref{ay}) we obtain the  relationship between $y$ and $z$. Namely,  
\begin{equation}
\label{z2}
z_{2}(y) = \left(\frac{1 + \bar{y}}{1 + y}\right)^{1/(1 - \beta)} - 1.
\end{equation}
For the early evolution described by (\ref{original a1}), using  (\ref{a1 in terms of x}) we obtain
\begin{equation}
\label{z1}
z_{1}(x) = \zeta\left(\frac{1 + x}{x}\right)^{1/(\alpha -1)} - 1,
\end{equation}
where 
\begin{equation}
\label{zeta}
\zeta = \left(\frac{{\stackrel{\ast}{x}}}{1 + {\stackrel{\ast}{x}}}\right)^{1/(\alpha -1)}\left(\frac{1 + \bar{y}}{1 + {\stackrel{\ast}{y}}}\right)^{1/(1 - \beta)},
\end{equation}
which guarantees that $z_{1}({\stackrel{\ast}{x}}) = z_{2}({\stackrel{\ast}{y}})$.

The above formulae allow us to translate the discussion from the theoretical $x$-and-$y$-language into the practical $z$-language.

\section{The era of transition}

In this section we discuss the ``thin" era sandwiched between, the end of the early evolution described by  $a_{1}$, and the beginning of the late evolution of the universe, as described by $a_{2}$. 

First we note that the dominant energy condition allows different parts of the universe to evolve, from one phase to another, at different values of $x$ (or $z$). Therefore, there is no reason to expect that  the transition from decelerated to accelerated cosmic expansion is an instantaneous
event, occurring simultaneously all over the universe.

This picture corresponds to the creation of bubbles of  new phase, in the middle of the old one, typical of first-order phase transitions. These bubbles expand and collide until the old phase disappears completely and the phase transition is complete.

Our formulae allow us to predict the epoch when the bubbles of accelerating phase start forming  and when the transition is complete. In fact, 
using (\ref{more stringent constraint on y}) and (\ref{z2}) we find

\begin{equation}
\label{the era of transition in terms of z}
{\stackrel{\ast}{z}}_{min} <  {\stackrel{\ast}{z}}  <  {\stackrel{\ast}{z}}_{max},
\end{equation}
where
\medskip
\begin{equation}
{\stackrel{\ast}{z}}_{min} = \left[\frac{(3\bar{\Omega}_{m} + 2\beta - 2)(\alpha - \beta)}{\alpha(3\bar{\Omega}_{m} - \beta)}\right]^{1/(1 - \beta)} - 1,
\end{equation}
and 
\begin{equation}
\label{zmax}
{\stackrel{\ast}{z}}_{max} = \left[\frac{[2\alpha(\alpha + 2) - \beta(3\alpha - 2)](3\bar{\Omega}_{m} + 2\beta - 2)}{2\alpha(2 + \alpha)(3\bar{\Omega}_{m} - \beta)}\right]^{1/(1 - \beta)} - 1.
\end{equation}
\medskip
This can also be expressed in terms of  the universal time. First, we note that  
the time $\stackrel{\ast}{t}$ at which the transition occurs is given by,
\begin{equation}
\stackrel{\ast}{t} = \bar{t}\left(\frac{\stackrel{\ast}{y}}{\bar{y}}\right)^{\beta/2(1 - \beta)},
\end{equation}
where $\bar{y}$ and $\bar{t}$ are provided  by (\ref{condition for beta}) and  (\ref{age of the universe}). Then, using (\ref{more stringent constraint on y}) we find that
\medskip

\begin{equation}
\label{the era of transition in terms of t}
{\stackrel{\ast}{t}}_{min} <  {\stackrel{\ast}{t}}  <  {\stackrel{\ast}{t}}_{max},
\end{equation}
with

\begin{equation}
\label{tmin}
{\stackrel{\ast}{t}}_{min} = \bar{t}\left[\frac{\beta(3\alpha - 2)(3\bar{\Omega_{m}} - \beta)}{(3\beta - 2)[2\alpha(2 + \alpha) - \beta(3\alpha - 2)]}\right]^{\beta/2(1 - \beta)},
\end{equation}
and
\medskip
\begin{equation}
\label{tmax}
{\stackrel{\ast}{t}}_{max} = \bar{t}\left[\frac{\beta(3\bar{\Omega}_{m} - \beta)}{(3\beta - 2)(\alpha - \beta)}\right]^{\beta/2(1 - \beta)}.
\end{equation}

\paragraph{Simultaneous transition:}

We note that for $\alpha = 6$, 
${\stackrel{\ast}{t}}_{max} = {\stackrel{\ast}{t}}_{min}$ and ${\stackrel{\ast}{z}}_{max} = {\stackrel{\ast}{z}}_{min}$. Which means that the transition occured at the same time, all over the universe (regardless of $\bar{\Omega}_{m}$ and the choice of $\beta$), if the very early universe was filled with stiff  matter $p = \rho$.

\subsection{The transition for different primordial scenarios}

The question we want to elucidate here is how the details of the transition depend on the equation of state of the primordial matter. Specifically, when does it start?, how long does it take?

The above formulae depend on three parameters, viz., $\alpha$, $\bar{\Omega}_{m}$ and $\beta$. 
Parameter $\alpha$ is related to the equation of state of the primordial matter in the  very early universe; $\alpha = 2, 3,4,6$ correspond to a network of cosmic strings; dust; radiation and stiff matter, respectively.

Measurements of the density parameter today, based on dynamical techniques, indicate that $\bar{\Omega}_{m} \approx 0.1 - 0.3$. Here, in order to get some explicit results, we take $\bar{\Omega}_{m} = 0.3$, which seems to be favored by observations.  

Regarding $\beta$, besides the constraint (\ref{range of beta}), 
the theory provides no mechanism, in terms of basic principles, to specify its exact value. However, an estimated value comes from the age of the universe. Indeed, from (\ref{age of the universe}), it follows that  
\begin{equation}
\label{equation for beta}
F(\bar{\Omega}_{m}, \beta) = \bar{t} \bar{H},
\end{equation}
which is an equation that allows us to get $\beta$ as soon as we provide $\bar{t}$, $\bar{H}$ and $\bar{\Omega}_{m}$.

Recent cosmological data from WMAP \cite{Spergel} indicate that the age of the universe is $13.7 \pm 0.2$ Gyr. If we take $\bar{t} = 13.7$Gyr and   $\bar{H} = 0.7 \times 10^{- 10}$yr$^{-1}$, then
\begin{equation}
\label{estimating beta}
\beta = 0.700.
\end{equation}
We will take this value in the examples bellow. 

\subsubsection{When does the transition start?}

Here  we restrict the discussion to $2 \leq  \alpha \leq 6$, because for $\alpha < 2$ the early universe is in a state of accelerated expansion from the beginning. 

The first conclusion from our equations  is that the transition started very ``recently". Namely,  it sets off at 
\begin{equation}
z = (1.03, \;\;1.13,\;\; 1.28,\;\; 1.55),\;\;\; t = (5.0, \;\;4.6, \;\;3.9,\;\; 2.9)\;Gyr,
\end{equation}
for $\alpha = (2,\;\; 3, \;\;4, \;\;6)$, 
respectively. We notice that recent in terms of $z$ has not the same meaning as recent in terms of $t$. For example, for stiff matter $z = 1.55$ 
is close $z = 0$, which is the present. But $t = 2.9$Gyr is far from $13.7$Gyr, which is the age of the universe.

The important result is that   
the transition  is ``almost" insensitive to the equation of state in the very early universe; it begins at approximately the same redshift era, viz., $z \sim 1-1.5$, for a wide variety of primordial matter.

\subsubsection{How long does it take?}

The  time-span $\delta t = ({\stackrel{\ast}{t}}_{max} - {\stackrel{\ast}{t}}_{min})$ taken for the  transition to complete decreases (increases) with the increase (decreases) of $\alpha$. Namely,
\begin{equation}
\delta t = (9.9, \;\;3.1,\;\; 1.1,\;\; 0)Gyr, 
\end{equation}
for $\alpha = (2,\;\; 3, \;\;4, \;\;6)$, 
respectively.

Clearly, the duration of the transition   
depend on how ``stiff",  or ``soft"  is the primordial matter; the softer the equation of state,  the longer the transition.

\medskip

Let us consider some examples in more detail. 

\paragraph{Dust:} For $\alpha = 3$ the transition starts when the universe is about $4.6$Gyr and terminates when it is about $7.7$Gyr. In terms of the redshift, the era of transition  extends over the interval  
\begin{equation}
{\stackrel{\ast}{z}} \in (0.59, 1.13). 
\end{equation}

\paragraph{Radiation:} For $\alpha = 4$ the transition starts when the universe is about $3.9$Gyr and terminates when it is about $5.0$Gyr. In terms of the redshift, the era of transition extends over the interval  
\begin{equation}
\label{narrow era}
{\stackrel{\ast}{z}} \in (1.03, 1.28). 
\end{equation}

\paragraph{Stiff matter:} As it was mentioned before, for $\alpha = 6$, the transition is an instantaneous phenomenon, which occurs everywhere at the same redshift. In the case under consideration,  ${\stackrel{\ast}{z}}_{min} = {\stackrel{\ast}{z}}_{max} \approx 1.55$. The transition starts earlier than for any other type of primordial mater

\subsection{The transition for primordial radiation} 

Let us consider in more detail the transition for  the case where the primordial universe consists of radiation and ultra-relativistic matter, which is the most  accepted scenario. 

For $\beta = 0.7$, we find  $\bar{y} = 0.5$ and $0.171 < {\stackrel{\ast}{y}} < 0.212$. Consequently, the  transition occurs in the narrow era given by (\ref{narrow era}), or
\begin{equation}
\label{allowed z}
{\stackrel{\ast}{z}} = 1.16 \pm 0.12.
\end{equation}
The dominant energy condition requires ${\stackrel{\ast}{x}} < x_{max} = 0.2$.
In principle, different parts of the universe can start the transition at different times\footnote{ 
From (\ref{a1 in terms of x}) it follows 
that $a \sim x ^{1/3}$, for $ \alpha = 4$ and small values of $x$, which means that  $x$ is proportional to the volume of the radiation-dominated early universe at the moment of transition.}; each  ${\stackrel{\ast}{x}} \in (0, 0.2)$ corresponds to a precise  $z$ within its range (\ref{allowed z}).

In order to see the fitting formulae at work, let us take ${\stackrel{\ast}{x}} = 0.1$. Then, from (\ref{x from the boundary conditions}) and (\ref{positivity of eta}) we obtain ${\stackrel{\ast}{y}} = 0.189$ and $\eta = 62.216$, respectively. This corresponds to 
\begin{equation}
{\stackrel{\ast}{z}} = 1.17,\;\;\;{\stackrel{\ast}{t}} = 0.32 \bar{t} \approx 2\mbox{Gyr},
\end{equation}

\begin{equation}
\label{the constants for primordial radiation}
C_{1} =  1.297 \bar{H}^{- 3/4},\;\;\;C_{2} =  0.722\bar{H}^{3/7},\;\;\;
\frac{A_{1}}{A_{2}}  =  3.962,
\end{equation}
and
\begin{equation}
\label{zeta in our example}
\zeta = 0.975
\end{equation} 

Let us now collect results and outline the evolution of the universe according to our model.

\medskip

As the early universe expands the deceleration parameter increases with time, viz.,
\begin{equation}
\label{q1 for the example}
q_{1} = 1 + 5x,\;\;\;x = C_{1}^{- 2}t^{3/2},\;\;\;x = \left[\left(\frac{z + 1}{\zeta}\right)^{3} - 1\right]^{- 1},
\end{equation}
where $\zeta$ and  $C_{1}$ are  given by (\ref{the constants for primordial radiation}) and (\ref{zeta in our example}), respectively. 
Note that the   largest observed redshift so far $z \sim 1100$, which is that of the CMB (the surface of last scattering), corresponds to $x = 0.695\times 10^{- 9}$.  So that  $q_{1} = 1$ in that era, as it should be for radiation. 
Then, it remains  positive throughout the whole early evolution, including the epoch  of reionization of the intergalactic medium, which occurred at $z \sim 11-6$ $(x \sim 5\times 10^{- 4} - 3\times 10^{- 3})$ \cite{Becker}-\cite{Kogut}.

 The early evolution described by (\ref{original a1}) ends by $z = 1.16 \pm 0.12$, when the universe is about $(4.4 \pm 0.5)$Gyr years old\footnote{We note that the specification of an era in terms of the physical time $t$ strongly depends on the details of the model, while $z$ depends only on the FRW nature of the metric and the assumption of flat geometry. Said another way, the same $z$ might corresponds to very different $t$'s, depending on the model.}. Then,   there is  a phase transition where the deceleration parameter changes from (\ref{q1 for the example}) to 
\begin{equation}
q_{2} = - 0.65 + \frac{0.05}{y},\;\;\;y = C_{2}^{2}t^{6/7}.
\end{equation}
For ${\stackrel{\ast}{x}} = 0.1$, the deceleration parameter jumps from $q_{1} = 1.5$ to $q_{2} = - 0.385$,  at $z = 1.17$. From then on, i.e., for $z < 1.17$ the universe is speeding up its expansion. 

What is important here is that  
the deceleration parameter today is $q _{2} = - 0.55$, regardless of the specific choice of ${\stackrel{\ast}{x}} \in (0, 0.2)$, which follows from (\ref{condition for beta}) for $\bar{\Omega}_{m} = 0.3$.

The densities $\rho_{1}$ and $\rho_{2}$  are continuous during the transition. However, the pressure is discontinuous.  To be precise, it is positive during the early evolution $(\infty < z < 1.16 \pm 0.12)$, viz.,   
\begin{equation}
8\pi G p_{1} = \frac{1 + 10 x}{4t^2(1 + x)^2},\;\;\;0 \leq x \leq 0.2,
\end{equation}
and is negative during the late evolution $(z < 1.16 \pm 0.12)$. Specifically,  
\begin{equation}
8\pi G p_{2} = \frac{0.4y - 9.2y^2}{0.49t^2(1 + y)^2},\;\;\;0.171 \leq y \leq 0.5. 
\end{equation}
We have already mentioned that a negative pressure can be associated with the energy of a slowly evolving cosmic scalar field. 
The paradigm is that the energy density and pressure of the universe can be separated in components; ordinary matter and a dark energy, which can be  modeled by  a scalar field. Namely,
\begin{eqnarray}
\label{splitting the effective quantities}
8\pi G\rho &=& 8\pi G\rho_{m} + \frac{1}{2}{\dot{\phi}}^2 + V(\phi),\nonumber \\ 
8\pi Gp &=& 8\pi Gp_{m} + \frac{1}{2}{\dot{\phi}}^2 - V(\phi).
\end{eqnarray}
At low redshifts the universe is matter dominated. Consequently,
\begin{equation}
8\pi G p = {\dot{\phi}}^2/2 - V(\phi).
\end{equation} 
for the most part of the evolution of the universe. Thus, the discontinuity in $p$ is produced by a discontinuity in the evolution of the scalar field. 

The evolution of $\phi$, which  is determined by the equation 
\begin{equation}
\label{field equation}
\ddot{\phi} + 3 \frac{\dot{a}}{a}\dot{\phi} + \frac{d V}{d \phi} = 0,
\end{equation}
has two qualitatively different regimes. The slow-evolving  regime corresponds to the late evolution of the universe, with negative pressure, where $V$ dominates over the kinetic term ${\dot{\phi}}^2/2$.  For $V \gg {\dot{\phi}}^2 $, the potential mimics a variable cosmological term, and the energy density of the universe splits up as  $\rho_{2} = \rho_{2m} + \Lambda/8\pi G$, with 
 \begin{equation}
8 \pi G \rho_{2m} = \frac{7y^2 + y}{1.225t^2(1 + y)^2}, \;\;\;
\Lambda = \frac{23y^2 - y}{1.225t^2(1 + y)^2}, \;\;\;0.171 \leq y \leq 0.5.
\end{equation}
 The rapidly-evolving  regime corresponds to the early evolution of the universe, with positive pressure, where the kinetic term ${\dot{\phi}}^2/2$ dominates over $V$.  For ${\dot{\phi}}^2\gg  V$, the field energy approximates the effect of a perfect fluid  with the equation of state $p_{\phi} = \rho_{\phi}$. For the case under consideration the energy density of the universe splits up as  $\rho_{1} = \rho_{1m} + \rho_{\phi}$, with 

 \begin{equation}
8 \pi G \rho_{1m} = \frac{1 - 5x}{2t^2(1 + x)^2}, \;\;\;8\pi G \rho_{\phi} = \frac{1 + 10x}{4t^2(1 + x)^2}, \;\;\;0 \leq x \leq 0.2.
\end{equation}
\subsection{Latent heat} 
We note that although the total energy density is continuous at the transition, the discontinuity in the scalar field induces a discontinuity in the density of ordinary matter. In our case, for the typical values $x = 0.1$ and $y = 0.189$,  we get $\rho_{2m} \approx 1.23\rho_{1m}$. 

This is perfectly consistent with the notion that first-order phase transitions involve a latent heat, which is either absorbed or released during the transition. 

In the present context, the latent heat is  the kinetic energy of the scalar field accumulated during the early evolution of the universe. 
Then, during the transition this energy is released; $88\%$ goes to the formation of the cosmological term, while the rest $12\%$ transforms into ordinary matter. 

\section{Discussion}

We have seen that the recent accelerated cosmic expansion appears as  a natural consequence of embedding our universe, as a dynamical four-dimensional  hypersurface, in the  five-dimensional bulk spacetime given by (\ref{Ponce de Leon solution}). This acceleration is an effect from the gravitational field in the bulk, which is transmitted to $4D$ through the dynamics of spacetime sections $\Sigma: y = f(t)$. 

We arrive at this point of view after calculating the effective matter density $\rho_{eff}$ induced on $\Sigma$, which can be separated in two parts \cite{JPdeL/0511067}
\begin{equation}
\rho_{eff} = \rho_{0} + \rho_{f}.
\end{equation}
where\footnote{Here $a = a(t, y)_{|y = f(t)}$ denotes the cosmic scale factor induced on spacetime sections $\Sigma: y = f(t)$ and $k = 0, + 1, -1$. Also $\dot{a}$ and $a'$ denote $\partial a/\partial t$ and $\partial a/\partial y$, both evaluated at $y = f(t)$, respectively \cite{JPdeL/0511067}. Thus, matter quantities depend on $t$, but {\it not} on $y$.}
\begin{eqnarray}
\label{matter density}
8 \pi G \rho_{0} &=& 3\left(\frac{\dot{a}}{a}\right)^2 + \frac{3 k}{a^2},\nonumber \\
8 \pi G \rho_{f} &=& 3\left[\frac{2 {\dot a}a'}{a^2}{\dot f} + \left(\frac{a'}{a}\right)^2 {\dot f}^2\right].
\end{eqnarray}
The first term $\rho_{0}$ is the usual general relativity term recovered on $\Sigma: y = y_{0}$ hypersurfaces, while the second one is a ``higher-dimensional" correction. Similarly, the effective pressure $p_{eff}$ can be separated as
\begin{equation}
p_{eff} = p_{0} + p_{f},
\end{equation}
with
\begin{eqnarray}
\label{matter pressure}
8 \pi G p_{0} = - \left(\frac{2 \ddot{a}}{a} + \frac{{\dot{a}}^2}{a^2}\right) - \frac{k}{a^2},
\end{eqnarray}
\begin{equation}
8 \pi G p_{f} = - \left[\left( \frac{4 {\dot{a}}'}{a}  
+ \frac{ 2\dot{a}a'}{a^2}\right)\dot{f} 
+ \left(\frac{2 a''}{a} + \frac{{a'}^2}{a^2}\right) {\dot{f}}^2
+ \left(\frac{2a'}{a}\right)\ddot{f}\right].
\end{equation}
Thus, the solutions of (\ref{defining f}) lead to two distinct situations. If $f = constant$, then $\rho_{f} = p_{f} = 0$ and the effective matter content  in $4D$ is {\it identical} to the one in four-dimensional general relativity, i.e., $\rho_{0}$ and $p_{0}$. 
For the dynamical embedding provided by the second solution (\ref{solution for f}), neither $\rho_{f}$ nor $ p_{f}$ vanish. Therefore, they  play a crucial role in the evolution of the  universe.  Indeed, $\rho_{f}$ and $p_{f}$ look as the density and pressure of some ``new" matter and represent  higher-dimensional corrections (due to the motion of the spacetime section $\Sigma :y = f(t)$ in the bulk) to the ordinary matter quantities in four-dimensional general relativity.   

From a four-dimensional point of view, not aware of the extra dimension,  these new matter terms are associated with some kind of ``dark" energy, which is  modeled in different ways, e.g., as a cosmological term or   quintessence, as we did at the end of section $5.2$.  In this context, the cross-over from decelerated to accelerated cosmic expansion can be interpreted as  a first-order phase transition associated with a  discontinuity in a  Higg's-type scalar field. Which  in turn might suggest that the underlying mechanism for the 
transition has to do with quantum effects\footnote{This leads to conjecture that there might be an  intricate connection between extra dimensions and quantum effects that, eventually,  can lead to phase transitions at the macroscopic level.}.

Regarding the model discussed here, it is important to  emphasize that, although  metrics (\ref{original a1}) and (\ref{original a2}) seem to be different, they represent the same spacetime in different parameterization. 

We also  note that the dominant energy condition plays a decisive role, for setting  the limits of applicability of these models, and also for giving the era of transition with great precision ($z = 1.16 \pm 0.12$ for primordial radiation).

In our model, the deceleration parameter as well as the equation of state of the universe depend only on the density parameter $\Omega_{m}$. We obtain
\begin{equation}
\bar{q} = (- 0.55, \; - 0.7,\;- 0.85), \;\;\;\bar{w} = (- 0.7, - 0.8, - 0.9)\;\; \mbox{for}\;\;\;\bar{\Omega}_{m} = (0.3,\;0.2,\;0.1),
\end{equation}
respectively. These results are consistent with observations and other studies. In particular, they are similar to those obtained for quintessence \cite{Efstathiou}, tracker fields \cite{Steinhard}, braneworld theory \cite{JPdeL0401026}-\cite{JPdeL0412005}, model independent analysis of the redshift from SNe and radio galaxies \cite{Moncy}-\cite{DD1}, and Friedmann thermodynamics \cite{Bayin2}.

It is remarkable that similar results  are obtained from very different approaches. If we use that $\Omega_{m} + \Omega_{\Lambda} = 1$, then our formulae for the deceleration parameter (\ref{general formula for the acceleration}) and the equation of state of the universe (\ref{equation of state for the universe}) can be written as 
\begin{equation}
q = 2 - \frac{3}{2}(\Omega_{m} + 2\Omega_{\Lambda}),
\end{equation}
and
\begin{equation}
w =  1 - \Omega_{m} - 2\Omega_{\Lambda}.
\end{equation}
These expressions are {\it identical} to those obtained in braneworld models with variable vacuum energy \cite{JPdeL0401026}, on a fixed hypersurface 
$\Sigma_{0}: y = y_{0}$. They reduce to the appropriate FRW-counterparts for $\Omega_{\Lambda} = 0$ and $\Omega_{m} = 1$. In particular we obtain $q = 1/2$ and $w = 0$ as in the dust-FRW cosmologies.


\begin{thebibliography}{99}
\bibitem{Riess}{A.G. Riess {\it et al.,}  Supernova Search Team Collaboration, {\em Astron. J.}, {\bf 116}, 1009 (1998),  astro-ph/9805201.}
\bibitem{Perlmutter}{S. Perlmutter {\it et al.,} Supernova Cosmology Project Collaboration,   {\em Astrophys. J.}, {\bf 517},
565 (1999), astro-ph/9812133.}

\bibitem{Liddle}{Andrew R Liddle,  {\em New Astron.Rev.}, {\bf 45}, 235(2001), astro-ph/0009491.}
\bibitem{Seto}{N. Seto, S. Kawamura and T. Nakamura,  {\em Phys.Rev.Lett.} {\bf 87}, 221103(2001), astro-ph/0108011.}

\bibitem{Knop}{R. A. Knop {\em et al},  {\em Astrophys. J.}, {\bf 598},  102(2003), astro-ph/0309368.}
\bibitem{Tonry}{J.L. Tonry et al., Astrophys. {\bf J. 594},  1 (2003), astro-ph/0305008.}

\bibitem{Lee}{A.T. Lee et al, {\em Astrophys. J.}, {\bf 561},  L1(2001), astro-ph/0104459.}


\bibitem{Stompor}{R. Stompor et al, {\em Astrophys. J.}, {\bf 561},  L7(2001), astro-ph/0105062.}
\bibitem{Halverson}{
N.W. Halverson et al,  {\em Astrophys. J.}, {\bf 568},  38(2002), astro-ph/0104489.}

\bibitem{Netterfielf}{C.B. Netterfield et al,   {\em Astrophys. J.}, {\bf 571}, 604(2002), astro-ph/0104460.}

\bibitem{Pryke}{C. Pryke, {\it et al.,}  {\em Astrophys. J.}, {\bf 568},  46(2002),  astro-ph/0104490.}
\bibitem{Spergel}{D.N. Spergel {\it et al.,} {\em Astrophys. J.Suppl.},  {\bf 148}, 175 (2003), astro-ph/0302209.}
\bibitem{Sievers}{J. L. Sievers, {\it et al.,}  {\em Astrophys. J.},  {\bf 591},  599(2003), astro-ph/0205387.}
\bibitem{Riess2}{A.G. Riess {\em et al}, {\em Astrophys.J.}, {\bf 560}, 49(2001), astro-ph/0104455.}
\bibitem{Turner3}{M. S. Turner and  A. Riess, {\em Astrophys.J.}, {\bf 569}, 18(2002), astro-ph/0106051.}
\bibitem{Peebles}{P. J. E. Peebles and  B. Ratra, {\em Rev.Mod.Phys.} {\bf 75}, 559(2003), astro-ph/0207347.}
\bibitem{Padmanabhan0}{T. Padmanabhan, {\em Phys.Rept.} {\bf 380},  235(2003), hep-th/0212290.}
\bibitem{Zlatev}{I. Zlatev, L Wang and P. J. Steinhardt, {\em Phys.Rev.Lett.} {\bf 82}, 896(1999), astro-ph/9807002}
\bibitem{Armendariz}{C. Armendariz, V. Mukhanov, P. J. Steinhardt, {\em Phys.Rev.Lett.} {\bf 85},  4438(2000), astro-ph/0004134.}
\bibitem{Caldwell1}{R.R. Caldwell, R. Dave, P. J. Steinhardt, {\em Phys.Rev.Lett.} {\em 80}, 1582(1998), astro-ph/9708069.}
\bibitem{Deustua}{S.E. Deustua, R. Caldwell, P. Garnavich, L. Hui, A. Refregier, 
``Cosmological Parameters, Dark Energy and Large Scale Structure", astro-ph/0207293.}

\bibitem{Krause}{A. Krause and Siew-Phang Ng, {\em Int.J.Mod.Phys.} {\bf A21} 1091(2006), hep-th/0409241.}
\bibitem{Caldwell2}{R. R. Caldwell, M. Kamionkowski, N. N. Weinberg, {\em Phys.Rev.Lett.} {\bf 91},  071301(2003),  astro-ph/0302506.}
\bibitem{Caldwell3}{R.R. Caldwell, {\em Phys.Lett.} {\bf B545}, 23(2002), astro-ph/9908168.}
\bibitem{Nojiri}{S. Nojiri and S. D. Odintsov, 
{\em Phys.Lett.} {\bf B562},  147(2003), hep-th/0303117; {\em Phys.Rev.} {\bf D68}, (2003) 123512, hep-th/0307288.}
\bibitem{Stefancic}{H. Stefancic, {\em Phys.Lett.},  {\bf B586},  5(2004), astro-ph/0310904; {\em Eur.Phys. J.},  {\bf C36},  523(2004),  
 astro-ph/0312484.} 
\bibitem{Turner1}{S. M. Carroll, V. Duvvuri, M. Trodden and M. S. Turner, {\em Phys.Rev.}, {\bf  D70},  043528(2004), astro-ph/0306438.}
\bibitem{Turner2}{G. Dvali and M. S. Turner, ``Dark Energy as a Modification of the Friedmann Equation", astro-ph/0301510.}
\bibitem{Bayin}{S. Bayin, {\em Int. J. of Mod. Phys.} {\bf D11}, 1523(2002), astro-ph/0211097.}
\bibitem{Mota}{D.F. Mota and J.D. Barrow,  {\em Mon. Not. Roy. Astron. Soc.},  {\bf 349} 281(2004), astro-ph/0309273; 
{\em Phys.Lett.} {\bf B581}, 141(2004), astro-ph/0306047.}
\bibitem{Yungui}{Yungui Gong and Chang-Kui Duan, {\em Class.Quant.Grav.},  {\bf 21},  3655(2004), gr-qc/0311060.}
\bibitem{Capozziello}{S. Capozziello, S. Carloni, A. Troisi, 
"Recent Research Developments in Astronomy and Astrophysics"-RSP/AA/21-2003,  astro-ph/0303041.}
\bibitem{Dolgov}{A.D. Dolgov, M. Kawasaki, 
{\em Phys.Lett.}, {\bf B573},  1(2003), astro-ph/0307285.}

\bibitem{Easson}{D. A. Easson, {\em Int.J.Mod.Phys.}{\bf A19},  5343(2004), astro-ph/0411209.}

\bibitem{Lue}{A. Lue, G. Starkman, {\em Phys.Rev.} {\bf D67}, 064002(2003), astro-ph/0212083.}
\bibitem{Deffayet1}{C. Deffayet, G. Dvali and  G. Gabadadze, {\em Phys.Rev.} {\bf D65} 044023(2002),  astro-ph/0105068.}
\bibitem{Deffayet2}{C. Deffayet, S. J. Landau, J. Raux, M. Zaldarriaga and P. Astier, {\em Phys.Rev.} {\bf D66},  024019(2002), astro-ph/0201164.}
\bibitem{Gorini}{V. Gorini, A. Kamenshchik and  U. Moschella, {\em Phys.Rev.} {\bf D67} 063509(2003),   
astro-ph/020939}
\bibitem{Neves}{R. Neves, C. Vaz, {\em Phys.Rev.} {\bf D68}, (2003) 024007, hep-th/0302030; {\em Phys.Lett.} {\bf B568},  153(2003), hep-th/0304266.}
\bibitem{Bento}{M.C. Bento, O. Bertolami, A.A. Sen, {\em Phys.Lett.} {\bf B575}, 172(2003), astro-ph/0303538; {\em Gen.Rel.Grav.} {\bf 35},  2063(2003), gr-qc/0305086; {\em Phys.Rev.}  {\bf D66},  043507(2002), gr-qc/0202064.}
\bibitem{Freese}{K. Freese and M. Lewis, {\em Phys.Lett.} {\bf B540} 1(2002), astro-ph/0201229.}
\bibitem{Gondolo}{P. Gondolo and  K. Freese, {\em Phys.Rev.} {\bf D68}, 063509(2003), hep-ph/0209322.}
\bibitem{JPdeL0305041}{J. Ponce de Leon, {\em Class.Quant.Grav.} {\bf 20}, 5321(2003), gr-qc/0305041.}
\bibitem{JPdeL0401026}{J. Ponce de Leon, {\em Gen.Rel.Grav.} {\bf 37},  53(2005), gr-qc/0401026.}
\bibitem{JPdeL0412005}{J. Ponce de Leon, {\em Gen.Rel.Grav.} {\bf 38},  61(2006), gr-qc/0412005.}
\bibitem{JPdeL 1}{J. Ponce de Leon, {\em Gen. Rel. Grav.} {\bf 20}, 539(1988).}

\bibitem{Overduin}{J.M. Overduin and P.S. Wesson, {\em Phys. Reports} {\bf 283}, 303(1997).}
\bibitem{Fukui Seahra and Wesson}{Takao Fukui, Sanjeev S. Seahra and P.S. Wesson, {\em J.Math.Phys.} {\bf 42}, 5195(2001), gr-qc/0105112. }
\bibitem{Billyard and Sajko}{A.P. Billyard and W.N. Sajko, {\em Gen.Rel.Grav.} {\bf 33}, 1929(2001), gr-qc/0105074.}
\bibitem{Wesson hd}{P.S. Wesson, {\em J.Math.Phys.} {\bf43}, 2423(2002), gr-qc/0105059.}
\bibitem{Inev of sing}{J. Ponce de Leon, {\em Mod. Phys. Lett.} {\bf A16}, No. 21, 1405(2001), gr-qc/0106020.}
\bibitem{Seahra 1}{S.S. Seahra, P.S. Wesson, {\em Class.Quant.Grav.} {\bf 19}, 1139(2002), gr-qc/0202010.}
\bibitem{Seahra 2}{P.S. Wesson and S.S. Seahra, {\em Ap. J.} {\bf L75}, 557(2001).}
\bibitem{equ. STM-Brane}{J. Ponce de Leon, {\em Mod. Phys. Lett.} {\bf A16}, No. 35, 2291(2001), gr-qc/0111011.}
\bibitem{Wesson book}{P.S. Wesson, {\em Space-Time-Matter} (World Scientific Publishing Co. Pte. Ltd. 1999).}



\bibitem{JPdeL/0511067}{J. Ponce de Leon, ``Reinventing spacetime on a dynamical hypersurface",  To appear in Mod. Phys. Lett. A, gr-qc/0511067.}

\bibitem{JPdeLgr-qc/0512067}{J. Ponce de Leon, ``Extra symmetry in the field equations in $5D$ with spatial spherical symmetry". To appear in Classical and Quantum Gravity, gr-qc/0512067.}


\bibitem{Moncy}{M. John, {\em Astrophys.J.}, {\bf 614}, 1(2004), astro-ph/0406444.}
\bibitem{DD1}{Ruth A. Daly and  S. G. Djorgovski, {\em Astrophys.J.}, {\bf 612}, 652(2004), astro-ph/0403664.}

\bibitem{Alam}{U. Alam, V. Sahni, T. Deep Saini and  A. A. Starobinsky, 
{\em Mon.Not.Roy.Astron.Soc.} {\bf 354}, 275(2004), astro-ph/0311364.}
\bibitem{Efstathiou}{G. Efstathiou, ``Constraining the equation of state of the Universe from Distant Type Ia Supernovae and Cosmic Microwave Background Anisotropies", astro-ph/9904356.}
\bibitem{Steinhard}{P.J. Steinhardt, L. Wang and I. Zlatev,  {\em Phys.Rev.} {\bf D59},  123504(1999),  astro-ph/9812313.}
\bibitem{Pelmutter2}{S. Perlmutter, M. S. Turner and M. White, {\em Phys.Rev.Lett.} {\bf 83}, 670(1999), astro-ph/9901052.} 
\bibitem{Becker}{R.H. Becker {\em et al}, {\em Astron. J.}, {\bf 122}, 2850(2001), astro-ph/0108097.}
\bibitem{Kogut}{A. Kogut {\em et al}, {Astrophys.J.Suppl.}, {\bf 148},  161(2003), astro-ph/0302213.}
\bibitem{Bayin2}{S. Bayin, ``Missing Mass, Dark Energy and the Acceleration of the Universe. Is Acceleration Here to Stay?", astro-ph/0410710.}




\end{thebibliography}
\end{document}